\documentclass[12pt,reqno]{article}
\usepackage{amsmath, amsthm, amssymb,epsfig,alltt}
\usepackage{mathtools}

\def\a{\alpha}

\def\p{\partial}
\def\m{\mu}

\def\t{\tau}

\def\s{\sigma}
\def\g{\gamma}

\def\half{\frac{1}{2}}

\def\barz{{\bar z}}

\def\sp{\sigma^\prime}
\def\nn{\nonumber}

\def\2pap{2\pi\alpha^\prime}

\def\beq{\begin{eqnarray}}
 \def\eeq{\end{eqnarray}}
 \def\4pap{4\pi\a^\prime}
 
 \def\sp{{\s^\prime}}

 \def\barz{{\bar z}}
 
 \def\barxi{{\bar \xi}}

 \def\barpsi{{\bar \psi}}

\begin{document}


\title{Chiral Fermion and Boundary State Formulation:
Resonant Point-Contact Tunneling}

\author{Taejin Lee \\~~\\
Department of Physics, Kangwon National University, \\
Chuncheon 200-701 Korea \\
email: taejin@kangwon.ac.kr}

\maketitle

\centerline{\bf Astract}

We study a model of resonant point-contact tunneling of a single Luttinger-liquid lead, 
using the boundary state formulation. The model is described by a single chiral Fermion 
in one dimensional space with one point contact interaction at the origin. It is the simplest model of 
this type. By folding the infinite line we map the model onto a non-chiral model on the half infinite line and transcribe the point-contact tunneling  interaction into a boundary interaction. The tunneling interaction 
yields a non-local effective action at the boundary. We explicitly construct the 
boundary state and evaluate the current correlation functions. The contact interaction is shown to act on 
the lower frequency modes more strongly.



\vskip 2cm

\section{INTRODUCTION}

The one dimensional system is an exciting arena where techniques and ideas from different fields 
influence each other. The one-dimensional model describing the point-contact tunneling \cite{Nayak}
between Luttinger-liquid \cite{Luttinger,Haldane} leads certainly belongs to this category. 
We discuss the model in its simplest form,
which has only a single chiral Fermion and a single lead. 
Despite its simple structure, the model exhibits its intriguing and essential
features of models of this type, including the Kondo model \cite{Hewson,Saleur1998,Saleur2000,Affleck:1990by} 
The model is defined on an infinite line with a point contact interactions at the origin.
This description of the model is called ``unfolded setup". The models can be also
studied in the ``folded setup" \cite{fendley1995a,fendley1995,wong,Affleck1994}, 
which is more suitable for 
field theoretical analysis. The chiral fields on the infinite line are mapped onto  
non-chiral fields on a half infinite line and the mid point contact interactions are transcribed 
into boundary interactions. The folded setup alludes that the boundary state formulation of string theory \cite{callan90,callan91,Callan1994,Lee:2005ge,Hassel,Lee:06,Lee2008,Lee2009} 
may be the most efficient framework to study the quantum impurity models of similar type. 
It is a simple matter to 
transcribe the models in the unfolded set up into the folded set up for the case of chiral boson models. 
However, it is not straightforward to transcribe the chiral Fermion models in the unfolded setup 
into the corresponding models in the folded setup. 

In this paper, we discuss details of the folding procedure to transcribe the chiral Fermion model into 
the unfolded setup. In the folded setup, the model is described by a Dirac spinor with two components.
Diving the infinite line into two half infinite lines, we get two boundaries. It is required 
that the boundary contact interaction term should be also extended appropriately. The Fermionic degrees of 
freedom of the impurity are represented by two-component spinors in the folded setup. 
Since the action for the spinors describing the impurity are only quadratic in Fermion fields, they can be
integrated out. As a result, we obtain a non-local effective action for the Fermi fields of the Luttinger liquid on the boundary, which is the world line of the contact point. In order to apply the boundary state formulation of string theory, we interchange $\t$ and $\s$. It brings us the closed string picture of the 
model. We construct the exact boundary state, which reproduces 
the continuity condition at the contact point in the unfolded setup. Using the boundary state, 
we calculate the correlation functions of the current operators. It is found that the contact tunneling interaction acts on the lower frequency modes more strongly. The high frequency modes are affected only weakly by the contact interaction. 

\section{Model: Chiral Fermion Model}

The one-dimensional model for the point-contact tunneling through a single Luttinger-liquid lead may be described by the following action of a single chiral Fermion field $\eta$ on an infinite line
\beq \label{chiralmodel}
S &=& \frac{1}{2\pi}\int^\infty_{-\infty} d\s \int d\t \Bigl\{ \eta^{\dag} \left(\p_\t + i\p_\s \right) 
\eta+ \delta(\s) d^{\dag} \p_\t d  + it \sqrt{2} \delta(\s)\left(\eta^{\dag} d+ \eta d^\dag \right) \Bigr\}.
\eeq
This model has been briefly discussed in the literature \cite{Nayak}. 
But the field theoretical analysis of the model has never been completed.
The Fermi fields $d$ and $d^\dag$ depict the degrees of freedom of impurity located at the origin.
They are the creation and annihilation operators of charge on the resonant state. 
In the action Eq.(\ref{chiralmodel}) we introduce a delta function in front of the kinetic term for the 
Fermi fields $d$ and $d^\dag$ so that the action for those Fermi fields is defined only on the world line of 
the contact point. If the delta function were not present, we would encounter a product of two delta functions when we evaluate the effective boundary action by integrating out $d$ and $d^\dag$. It yields
a divergence, which requires the renormalization of the coupling constant $t$. Here we avoid this 
unnecessary complication by simply defining the entire action for $d$ and $d^\dag$ only on the 
worldline of the contact point. The action can be given also in terms of a chiral boson field. But for the purpose of field theoretical calculation the Fermion action is more preferable, 
since the interaction term is only quadratic in Fermi fields. 

\subsection{Chiral Fermion model in the Unfolded Setup}

We may rewrite the action Eq.(\ref{chiralmodel}) in the following form by dividing it into the bulk action 
and the boundary action
\beq \label{chiralmodel2}
S 
&=& \frac{1}{2\pi}\int^\infty_{-\infty} d\s \int d\t \Bigl\{ \eta^{\dag} \left(\p_\t + i\p_\s \right) 
\eta \Bigr\} \nn\\
&&+ \frac{1}{2\pi} \int d\t\Bigl\{ d^{\dag} \p_\t d  + it\sqrt{2} \left(\eta^{\dag} d+ \eta d^\dag \right)\Bigr\}\Bigl\vert_{\s=0}.
\eeq
The action is defined on a two dimensional Euclidean space and all Fermi fields are anti-periodic in $\t$.
Thus, we are working on the theory at finite temperature. For the sake of simplicity we also scale
some physical variables appropriately in such a way that the range of $\t$ is $[0,2\pi]$. 
The equations of motion, which follow from the action Eq.(\ref{chiralmodel2}) are given by
\beq
(\p_\t + i\p_\s) \eta + it \sqrt{2}\delta(\s) d = 0, ~~ \p_\t d -it \sqrt{2} \eta \bigl\vert_{\s =0} = 0.
\eeq
Since the Fermi fields are anti-periodic in $\t$ we may expand the Fermi fields in terms of normal modes
\begin{subequations}
\beq
\eta(\t,\s) &=& \sum_n \eta_n(\s) e^{in\t}, ~~ d(\t) = \sum_n d_n e^{in\t}, \\
\eta^\dag(\t,\s) &=& \sum_n \eta_n(\s) e^{-in\t}, ~~ d^\dag(\t) = \sum_n d_n e^{-in\t}~~ n \in {\bf Z} + 1/2 .
\eeq
\end{subequations}
If we rewrite the equations of motion in terms of normal modes,
\begin{subequations}
\beq
n \eta_n  &=& -  \p_\s \eta_n  -t\sqrt{2}\delta(\s) d_n  \label{first1}\\
n d_n  &=& t \sqrt{2} \eta_n(0) = \frac{t}{\sqrt{2}} 
\left[\eta_n(0+)+ \eta_n(0-)\right] \label{second1}.
\eeq
\end{subequations}
For Eq.(\ref{first1}) to be defined continuously across the origin, the following condition must 
be imposed
\beq
- \left[\eta_n(0+) - \eta_n(0-)\right] - t \sqrt{2} d_n &=& 0. \label{normal3}
\eeq
Then it follows from Eqs.(\ref{second1},\ref{normal3}), 
\beq \label{continuity}
\eta_n(0+) = \left(\frac{1 - \frac{t^2}{n}}{1+\frac{t^2}{n}}\right)\, \eta_n (0-).
\eeq
This condition should be understood as the continuity condition at the contact point, which 
as we shall see shortly, corresponds to the boundary condition in the folded setup.

\subsection{Chiral Fermion Model in the Folded Setup}

The folding procedure begins with splitting the infinite one dimensional space into two half infinite 
lines. We rename the Fermi fields on the second half line as $\zeta$ and $c$ to distinguish them from the 
Fermi fields $\eta$ and $d$ on the first half line
\beq
S &=& \frac{1}{2\pi}\int^\infty_{0} d\s \int d\t \left[ \eta^{\dag} \left(\p_\t +i\p_\s \right) \eta
+ \delta(\s)d^{\dag} \p_\t d \right] \nn\\
&&+ \frac{1}{2\pi}\int^0_{-\infty} d\s \int d\t \left[ \zeta^\dag \left(\p_\t +i \p_\s \right) \zeta
+ \delta(\s)c^{\dag} \p_\t c \right]\\
&&+\frac{it \sqrt{2}}{2\pi}\int_{\s=0} d\t \left(\eta^{\dag} d+ \eta d^\dag \right). \nn
\eeq
For the time being we adopt the same boundary action as in the unfolded setup.  
The differential operator in the action, $\p$ should be understood as $\half( \overrightarrow{\p} - \overleftarrow{\p})$ \cite{brown}. Thus action for the 
Fermi field on the second half line may be written also as
\beq \label{equiv}
\frac{1}{2\pi}\int^0_{-\infty} d\s \int d\t \left[ \zeta^\dag \left(\p_\t +i \p_\s \right) \zeta\right] =
\frac{1}{2\pi}\int^0_{-\infty} d\s \int d\t \left[ \zeta \left(\p_\t +i \p_\s \right) \zeta^\dag\right].
\eeq
Splitting the infinite line into two half infinite lines, we create two boundaries. So appropriate boundary
condition should be imposed on the boundaries of the two half infinite lines. 
It follows from Eq.(\ref{equiv}) that we can choose one of the following two conditions
\beq
\eta(\t,0) = \zeta(\t,0)e^{i\theta}, ~~ \eta^\dag (\t,0) = \zeta^\dag(\t,0)e^{-i\theta}
\eeq
or 
\beq \label{second}
\eta(\t,0) = \zeta^\dag(\t,0)e^{i\theta},~~\eta^\dag(\t,0) = \zeta(\t,0)e^{-i\theta},
\eeq
as a boundary condition at $\s=0$. Here $\theta$ is a constant phase to be fixed later. 
We choose the second one Eq.(\ref{second}) as the boundary condition for a reason to be clear shortly. 

Then we take the parity transformation $\s \rightarrow -\s$ on the second half line to merge two
bulk actions into one. Exchanging the world sheet coordinates
$\t \rightarrow \s$ and $\s \rightarrow \t$ we get a closed string picture of the model
\beq \label{nonchiral}
S &=& \frac{1}{2\pi}\int^\infty_{0} d\t \int d\s \left[ i\eta^{\dag} \left(\p_\t -i\p_\s \right) \eta
+  \delta(\s) d^{\dag} \p_\s d \right] \nn\\
&&+ \frac{1}{2\pi}\int^\infty_{0} d\t \int d\s \left[ -i\zeta^{\dag} \left(\p_\t +i \p_\s \right) \zeta
+ \delta(\s) c^{\dag} \p_\s c \right]\\
&&+\frac{it\sqrt{2}}{2\pi}\int_{\t=0} d\s \left(\eta^{\dag} d+ \eta d^\dag \right) . \nn
\eeq
At the same time we shift the phases of the Fermi fields as 
\beq \label{phase}
\eta^\dag \rightarrow -i \eta^\dag, ~~\zeta^\dag \rightarrow i \zeta^\dag, ~~d^\dag \rightarrow -id^\dag,
~~c^\dag \rightarrow ic^\dag,
\eeq
to cast the bulk action into the familiar action of the Dirac spinor with two components
\beq
S &=& \frac{1}{2\pi}\int^\infty_{0} d\t \int d\s 
\left[\eta^{\dag} \left(\p_\t -i\p_\s \right) \eta+ \zeta^\dag\left(\p_\t +i \p_\s \right) \zeta \right] \nn\\
&&+\frac{1}{2\pi}\int_{\t=0} d\s \left[-id^\dag \p_\s d + ic^\dag \p_\s c + t\sqrt{2}\left(\eta^\dag d
+ \eta d^\dag\right) \right].\nn
\eeq
The path integral measure $D[\eta,\eta^\dag,\zeta,\zeta^\dag]$ for the Fermi fields is invariant under 
this phase shift. From the closed string action Eq.(\ref{nonchiral}) it is clear that $\eta$ is the right mover (anti-holomorphic function) and $\zeta$ is 
the left mover (holomorphic function)
\begin{subequations}
\beq
\zeta(\t,\s) &=& \psi_L(\t,\s),~~\zeta^\dag(\t,\s) = \psi_L^\dag(\t,\s),\\ 
\eta(\t,\s) &=& \psi_R(\t,\s),~~\eta^\dag(\t,\s) = \psi^\dag_R(\t,\s).
\eeq
\end{subequations}
The boundary condition Eq.(\ref{second}) is now read as 
\beq \label{second2}
\eta(\t,0) = i\zeta^\dag(\t,0)e^{i\theta},~~\eta^\dag(\t,0) = i\zeta(\t,0)e^{-i\theta} .
\eeq
If we choose $e^{i\theta} = e^{-i\theta} = -1$, this condition Eq.(\ref{second2}) can be 
indentified as the Neumann condition
\beq \label{neumann1}
\psi_L(\t,0) = i \psi^\dag_R(\t,0), ~~ \psi^\dag_L(\t,0) = i \psi_R(\t,0)
\eeq
(Note that we may represent the Fermi field operators $\psi_L$ and $\psi_R$ in terms 
of the boson fields $\phi_L$ and $\phi_R$ as 
\beq
\psi_L 
= e^{-\frac{\pi}{2} i \left(p_L + p_R \right)} e^{-\sqrt{2} i \phi_L}, ~~
\psi_R 
= e^{-\frac{\pi}{2} i \left(
p_L + p_R \right)} e^{\sqrt{2} i \phi_R}.
\eeq
In this representation the Neumann condition for the boson fields, $\phi_L = \phi_R$ can be 
transcribed into Eq.(\ref{neumann1}) for the Fermi fields $\psi$.)

Now the bulk action 
can be written succinctly in terms of two component Dirac spinor $\psi$
\beq
S[\psi] &=& \frac{1}{2\pi}\int^\infty_{0} d\t \int d\s \, \barpsi \g \cdot \p \,\psi .
\eeq
where $\g^0 = \s_1, ~~ \g^1 = \s_2, ~~ \g^5 = \s_3 = -i \g^0 \g^1 $ and
\beq
\psi = \begin{pmatrix} \psi_L \\ \psi_R \end{pmatrix}=\begin{pmatrix} \zeta \\ \eta \end{pmatrix}.
\eeq

As aforementioned the boundary interaction term should be also extended to be consistent 
with the folded setup  
\beq \label{twoterms}
t\sqrt{2} \left(\eta^\dag d + \eta d^\dag \right)  \rightarrow 
t \left(\eta^\dag d + \eta d^\dag + c^\dag \zeta +c \zeta^\dag\right) = i t\left(\barpsi \g^1 \xi
- \barxi \g^1 \psi\right) 
\eeq
where we define the two-component spinor $\xi$, representing the degrees of freedom of the impurity as
\beq
\xi = \begin{pmatrix} \xi_L \\ \xi_R \end{pmatrix}=\begin{pmatrix} c \\ d \end{pmatrix} .
\eeq
Under the Neumann condition, which may be extended to include the Fermi fields $\xi$
\begin{subequations}
\beq
\psi_L &=& i \psi^\dag_R, ~~ \psi^\dag_L = i \psi_R, \\
\xi_L &=& i \xi^\dag_R, ~~ \xi^\dag_L = i \xi_R ,
\eeq
\end{subequations}
the two interaction terms of the boundary action Eq.(\ref{twoterms}) are equivalent to each other. 
The coupling constant $t$ is scaled also in such a way that the folded model produces the same boundary condition as the unfolded model. Completing the folding procedure brings us the 
following action in the folded setup, to which the boundary state formulation is readily applicable 
\beq
S = \frac{1}{2\pi}\int^\infty_{0} d\t \int d\s \, \barpsi \g \cdot \p \,\psi + \frac{1}{2\pi} \int_{\t=0} d\s
\left\{ \barxi \g^1 \p_\s \xi + it \left(\barpsi \g^1 \xi - \barxi \g^1 \psi\right) \right\}.
\eeq

\section{Boundary State}

The advantages of the action in the folded setup are evident: Since model is depicted by the free closed 
string bulk action and the boundary action, which produces only boundary (initial) condition 
at the initial time $\t=0$, the free closed string Hamiltonian can be employed to evaluate 
the correlation functions of physical operators. We are ready to apply the boundary state formulation, 
which may be the most efficient method to 
calculate the partition function and the correlation functions of various physical operators.

We note that the action is only quadratic in the Fermi field $\xi$. Integrating out $\xi$ yields
a boundary effective action for $\psi$. It is convenient to make use of the mode expansions on a cylinder
\begin{subequations}
\beq
\psi (\t,\s) &=& \sum_n \psi_n(\t) e^{in\s}, ~~ \barpsi (\t,\s) = \sum_n \barpsi_n(\t) e^{in\s}, \label{normal1}\\
\xi (\s) &=& \sum_n \xi_n e^{in\s}, ~~ \barxi (\s) = \sum_n \bar{\xi}_n e^{in\s} \label{normal2}
\eeq
\end{subequations}
to explicitly evaluate the effective boundary action. 
For the Fermi fields on a cylinder, two conditions are available; the anti-periodic condition
\beq
\psi (2\pi) = -\psi (0), ~~ \barpsi (2\pi) = -\barpsi (0),~~
\xi (2\pi) = -\xi (0), ~~ \barxi (2\pi) = -\barxi (0),
\eeq
and the periodic condition
\beq
\psi (2\pi) = \psi (0), ~~ \barpsi (2\pi) = \barpsi (0),~~
\xi (2\pi) = \xi (0), ~~ \barxi (2\pi) = \barxi (0).
\eeq
The former is called the Neveu-Schwarz (NS) sector where the normal modes are labeled by half-odd-integers,
$n \in {\bf Z} + 1/2$
and the latter is called the Ramond (R) sector where the normal modes are labeled by integers, 
$n \in {\bf Z}$. Here we choose the NS sector only, because the vacuum in the R sector carries a
non-vanishing momentum while the vacuum in the NS sector does not. 
The physical vacuum belongs to the NS sector. 

\subsection{Effective Boundary Action}

If we rewrite the action for the Fermi field $\xi$ in terms of normal modes Eq.(\ref{normal1},\ref{normal2})
\beq
S[\barxi,\xi] = \frac{1}{2\pi} \int_{\t=0} d\s
\left\{ \barxi \g^1 \p_\s \xi + it \left(\barpsi \g^1 \xi - \barxi \g^1 \psi\right) \right\},
\eeq
we have 
\beq
S[\bar{\xi},\xi]=  i\int^\infty_0 d\t \sum_n   \left\{
n \barxi_{-n} \g^1 \xi_n + t 
\left( \barpsi_{-n} \g^1 \xi_n - \barxi_{-n} \g^1 \psi_n \right) \right\}.
\eeq
If we integrate out the Fermi field $\xi$, we obtain the effective action for $\psi$
\beq
\int D[\barxi,\xi]  \exp \left\{ -S[\bar{\xi},\xi] \right\} = \exp \left\{
- it^2 \sum_n \frac{1}{n} \barpsi_{-n} \g^1 \psi_n \right\} \Biggr\vert_{\t = 0}.
\eeq
This boundary effective action is non-local and quadratic in the Fermi field $\psi$. 
Note that we would encounter a divergence in the effective action in the form of product of 
two delta functions $(\delta(\t))^2 = \delta(\t) \delta(0)$ if
the delta function were not present in front of the kinetic term for the Fermi fields $d$ and $d^\dag$ in 
Eq.(\ref{chiralmodel}).

The boundary effective action $S_{\rm boundary}[\psi]$ may be written as 
\beq
S_{\rm boundary} = t^2 \int \frac{d\s}{2\pi} \barpsi \g^1 [\p_\s]^{-1} \psi \Bigr\vert_{\t=0},
\eeq
and it leads us to a formal expression of the boundary state 
\beq \label{formal}
\vert B \rangle = \exp \left[t^2 \int \frac{d\s}{2\pi} \barpsi \g^1 [\p_\s]^{-1} \psi \Bigr\vert_{\t=0}\right]
\vert N \rangle
\eeq
where $\vert N \rangle$ is the Neumann boundary state for the Fermi fields $\psi$. Correlation functions of the 
physical operators ${\cal O}_i$, $i=1, \dots, n$ may be evaluated as 
\beq
\langle {\cal O}_1 \cdots {\cal O}_n \rangle=
\langle 0 \vert :{\cal O}_1 \cdots {\cal O}_n: \vert B\rangle/\langle 0 \vert B\rangle.
\eeq

\subsection{The Exact Boundary State}

The explicit expression of the effective boundary action and the boundary state $\vert B\rangle$
can be obtained if the two spinor components of the Fermi fields, $\psi_L$ and $\psi_R$ are expanded 
in the Fermion oscillators
\begin{subequations}
 \label{generallabel} \begin{eqnarray} \psi_L(\tau+i\sigma)= \sum_n
 \psi_n e^{-n(\tau+i\sigma)} ~~,~~
 \psi^{\dagger}_L(\tau+i\sigma)=\sum_n
 \psi_n^{\dagger}e^{-n(\tau+i\sigma)} \label{fermode:a}\\
 \psi_R(\tau-i\sigma)=\sum_n \tilde \psi_n e^{-n(\tau-i\sigma)} ~~,~~
 \psi^{\dagger}_R(\tau-i\sigma) = \sum_n \tilde \psi_n^{\dagger}
 e^{-n(\tau-i\sigma)}.\label{fermode:b} \end{eqnarray}
 \end{subequations} 
The non-vanishing anticommutation relations between the Fermion operators are 
 \begin{equation} \label{anticomns} \left\{\psi_m, \psi_n^{\dagger}
 \right\} = \delta_{m+n} ~~~,~~~ \left\{ \tilde \psi_m , \tilde
 \psi_n^{\dagger}\right\}=\delta_{m+n}. 
 \end{equation} 
Since the Fermi fields $\psi_{L/R}$ are anti-periodic in the NS sector on a cylinder
the oscillators $\psi_n$ are labeled
by half-odd-integers, $n\in {\bf Z} +1/2$.
 
The oscillators can be uniquely separated into those with positive and negative modes. 
The vacuum in the NS sector is unique in contrast to the vacuum in the Ramond sector, which 
possesses Fermion zero mode oscillators.  
A representation of (\ref{anticomns}) is found by beginning with the vacuum state which
is annihilated by all positively moded oscillators 
\begin{equation}
\left.\begin{matrix} \psi_n \left| 0\right\rangle_{NS}=0,\,\, & \tilde
\psi_n\left|0\right\rangle_{NS}=0 \cr \psi^{\dagger}_n \left|
0\right\rangle_{NS}=0,\,\, & \tilde \psi^{\dagger}_n\left|0\right\rangle_{NS}=0
\cr \end{matrix} \right\}~~n>0 .\end{equation}
The Neumann state satisfies \cite{Lee:2005ge}
 \begin{subequations}
 \label{generallabel}
  \begin{eqnarray}
  \psi_L(0,\sigma)~\left|N \right\rangle &=&
  i\psi^{\dagger}_R(0,\sigma) \left|N \right\rangle ,\\
 \psi_L^{\dagger}(0,\sigma) \left|N \right\rangle &=&
  i \psi_R(0,\sigma)~\left| N\right\rangle ,
 \end{eqnarray}
 \end{subequations}
which are read in terms of the normal oscillators as
\begin{subequations}
\beq
\psi_n \vert N \rangle &=& i \tilde\psi^\dag_{-n} \vert N\rangle,~~
\psi^\dag_n \vert N \rangle = i \tilde \psi_{-n} \vert N\rangle, \label{normalaa} \\
\tilde\psi_n \vert N\rangle &=& -i \psi^\dag_{-n} \vert N \rangle,~~
\tilde\psi^\dag_n \vert N \rangle = -i \psi_{-n} \vert N\rangle  \label{normalbb}
\eeq
\end{subequations}
where $n$ is a positive half-odd-integer; $n \in {\bf Z}+ 1/2$, $n >0$.
A solution of these equations  in the NS sector is
 \beq
 \left| N\right\rangle =  :\exp\left\{\sum_{n=1/2}^\infty i
 \left(\psi_{-n}^{\dagger}\tilde \psi^{\dagger}_{-n}+ \psi_{-n}\tilde
 \psi_{-n} \right)\right\}: \vert 0 \rangle, \\
 n \in {\bf Z}+ 1/2,\quad n >0 .\nn
 \eeq
An explicit expression of the boundary state $\vert B\rangle$ follows from Eqs.(\ref{formal},\ref{fermode:a},
\ref{fermode:b}) 
\beq \label{explicit}
\vert B\rangle 
&=& :\exp \left\{- it^2 \int \frac{d\s}{2\pi} \left[\tilde\psi^\dag[\p_\s]^{-1} \tilde{\psi} 
 -\psi^\dag [\p_\s]^{-1} \psi \right]\right\} :\vert N \rangle \nn\\
&=& :\exp \left\{-t^2 \sum_n \frac{1}{n} \left( \tilde\psi^\dag_{-n} \tilde\psi_n + \psi^\dag_{-n}\psi_n\right)\right\} :
 \vert N \rangle \nn \\
&=& \prod_{n=1/2} :\exp \left\{-\frac{t^2}{n} \tilde\psi^\dag_{-n} \tilde\psi_n \right\}:
:\exp \left\{\frac{t^2}{n} \tilde\psi^\dag_{n} \tilde\psi_{-n} \right\}:\nn\\
&& 
:\exp\left\{-\frac{t^2}{n}\psi^\dag_{-n}\psi_n\right\} :
:\exp\left\{\frac{t^2}{n}\psi^\dag_{n}\psi_{-n}\right\} :
\vert N \rangle .
\eeq
 

We must check whether the constructed boundary state Eq.(\ref{explicit}) correctly reproduces the continuity condition Eq.(\ref{continuity}) of the model in the unfolded setup. 
It can be accomplished by rewriting the boundary state $\vert B\rangle$ as 
\beq \label{boundary3}
\vert B\rangle 
&=&\prod_{n=1/2}:\exp\left\{-\frac{t^2}{n} \psi^\dag_{-n} \psi_n \right\} \exp \left\{-\frac{t^2}{n} \tilde\psi_{-n} \tilde\psi_n^\dag \right\}: \nn\\
&& ~~\exp\left\{i\left(1-\frac{t^2}{n}\right) \psi^\dag_{-n} \tilde\psi^\dag_{-n}
\right\} \exp\left\{i\left(1-\frac{t^2}{n}\right) \psi_{-n} \tilde\psi_{-n} \right\}\vert 0 \rangle .
\eeq
Here we make use of the Neumann boundary condition Eqs.(\ref{normalaa},\ref{normalbb}). 
If we apply $\psi_n$, $n>0$ on the boundary state $\vert B\rangle$ Eq.(\ref{boundary3}), we find
\beq 
\psi_n \vert B\rangle = 
- \frac{t^2}{n} \psi_n \vert B\rangle + i \left(1- \frac{t^2}{n} \right) \tilde\psi^\dag_{-n} \vert B \rangle.
\eeq
Rearranging this equation we confirm that the obtained boundary condition 
coincides with the continuity condition in the unfolded setup 
\beq \label{boundary1}
\psi_n \vert B\rangle = i \left(\frac{1-\frac{t^2}{n}}{1+\frac{t^2}{n}}\right)\, \tilde\psi^\dag_{-n} \vert B\rangle.
\eeq
The phase factor $i$ on the RHS of Eq.(\ref{boundary1}) is due the phase shift Eq.(\ref{phase}) of the 
Fermi fields.

We can also check if other boundary conditions are correctly reproduced by applying oscillator operators $\psi^\dag_n$, $\tilde\psi_n$, $\tilde\psi^\dag_n$, $n>0$ on the boundary state $\vert B\rangle$. 
Rewriting the boundary state $\vert B\rangle$ as follows, using the Neumann condition Eqs.(\ref{normal1}
,\ref{normal2}) again, 
\beq
\vert B\rangle 
&=& \prod_{n=1/2} :\exp \left\{-\frac{t^2}{n} \tilde\psi^\dag_{-n} \tilde\psi_n \right\}:
:\exp \left\{-\frac{t^2}{n} \tilde\psi_{-n} \tilde\psi^\dag_{n}\right\}:\\
&& 
:\exp\left\{i\left(1-\frac{t^2}{n}\right)\psi^\dag_{-n}\tilde\psi^\dag_{-n}\right\} :
:\exp\left\{i\left(1-\frac{t^2}{n}\right)\psi_{-n}\tilde\psi_{-n}\right\} :
\vert 0 \rangle \nn
\eeq
we find 
\beq
\tilde\psi_n \vert B \rangle = -\frac{t^2}{n} \tilde\psi_n \vert B\rangle - i \left(1-\frac{t^2}{n} \right)
\psi^\dag_{-n} \vert B\rangle .
\eeq
This condition is equivalent to 
\beq \label{boundary22}
\tilde\psi_n \vert B \rangle = - i \left(\frac{1-\frac{t^2}{n}}{1+\frac{t^2}{n}}\right)\, \psi^\dag_{-n} \vert B\rangle .
\eeq
Similarly, if we recast the boundary state into the following form 
\beq
\vert B\rangle 
&=& \prod_{n=1/2} :\exp \left\{-\frac{t^2}{n} \tilde\psi^\dag_{-n} \tilde\psi_n \right\}:
:\exp \left\{-\frac{t^2}{n}\psi_{-n} \psi^\dag_{n} \right\}: \nn\\
&&
:\exp\left\{i\left(1-\frac{t^2}{n}\right)\psi_{-n}\tilde\psi_{-n}\right\} :
:\exp\left\{i\left(1-\frac{t^2}{n}\right)\psi^\dag_{-n}\tilde\psi^\dag_{-n}\right\} :
\vert 0 \rangle ,
\eeq
we find that the boundary state satisfies the following condition
\beq
\psi^\dag_n \vert B\rangle = - \frac{t^2}{n} \psi^\dag_n \vert B\rangle + i\left(1- \frac{t^2}{n} \right) \tilde\psi_{-n} \vert B\rangle ,
\eeq
which is equivalent to
\beq \label{boundary33}
\psi^\dag_n \vert B\rangle =  i \left(\frac{1-\frac{t^2}{n}}{1+\frac{t^2}{n}}\right)\, \tilde\psi_{-n} \vert B\rangle.
\eeq

Finally we may rewrite the boundary state $\vert B\rangle$ as follows
\beq
\vert B\rangle 
&=& \prod_{n=1/2}
:\exp \left\{-\frac{t^2}{n} \tilde\psi_{-n} \tilde\psi^\dag_{n} \right\}:
:\exp \left\{-\frac{t^2}{n} \tilde\psi^\dag_{-n} \tilde\psi_n \right\}:
 \nn\\
&&
:\exp\left\{i\left(1-\frac{t^2}{n}\right)\psi_{-n}\tilde\psi_{-n}\right\} :
:\exp\left\{i\left(1-\frac{t^2}{n}\right)\psi^\dag_{-n}\tilde\psi^\dag_{-n}\right\} :
\vert 0 \rangle .
\eeq
If we apply $\tilde\psi^\dag_{n}$ on the boundary state, we get 
\beq
\tilde\psi^\dag_{n} \vert B \rangle = - \frac{t^2}{n} \tilde\psi^\dag_n \vert B \rangle - 
i\left(1- \frac{t^2}{n} \right) \psi_{-n} \vert B\rangle.
\eeq
This condition can be rearranged as 
\beq \label{boundary44}
\tilde\psi^\dag_{n} \vert B \rangle 
= -i\left(\frac{1- \frac{t^2}{n}}{1+\frac{t^2}{n}}\right) \psi_{-n} \vert B\rangle.
\eeq
By some explicit algebra, we confirmed that the constructed boundary state in the folded setup
correctly reproduces the continuity condition in the unfolded setup.

As $t$ increases from zero to infinity, the conditions Eqs.(\ref{boundary1},\ref{boundary22},\ref{boundary33},\ref{boundary44}) interpolate from the Neumann boundary condition 
Eqs.(\ref{normalaa},\ref{normalbb}) to the Dirichlet boundary condition
\begin{subequations}
\beq 
\psi_n \vert D\rangle &=& -i \tilde\psi^\dag_{-n}  \vert D\rangle, ~~
\psi^\dag_n \vert D\rangle = -i \tilde\psi_{-n} \vert D \rangle, \label{Dirichleta}\\
\tilde\psi_n \vert D \rangle &=& i \psi^\dag_{-n} \vert D \rangle, ~~
\tilde\psi^\dag_{n} \vert D \rangle = i \psi_{-n} \vert D \rangle. \label{Dirichletb}
\eeq
\end{subequations}

\section{Correlation Functions of Current Operators}

Once we construct the boundary state it becomes a simple matter to calculate the correlation functions
of the operators, ${\cal O}_i$, $i= 1, \cdots, n$,
\beq
\langle T {\cal O}_1 \cdots {\cal O}_n \rangle &=& \langle 0 \vert :{\cal O}_1 \cdots {\cal O}_n: \vert B\rangle/ \langle 0 \vert B\rangle \nn\\
&=&  \langle 0 \vert :{\cal O}_1 \cdots {\cal O}_n: \exp\left(-S_{\text{boundary}}\right) \vert N\rangle /
\langle 0 \vert B\rangle.
\eeq
where $T$ denotes the $\t$ ordered product and $\langle 0 \vert$ is the vacuum state in the NS sector.  
Þgeneracy. So we should define the correlation functions of the operators as follows

As an example, we will calculate the correlation functions of the current operators
\beq
J_L = \psi^\dag_L \psi_L = \half(J^0 -iJ^1), ~~
J_R = \psi^\dag_R \psi_R = \half(J^0 +iJ^1),
\eeq
where $J^\m = \barpsi \g^\m \psi = \psi^\dag \g^0 \g^\m \psi$.
Suppose that a point particle moves to the left (right) toward the contact point. 
Then after the contact tunneling interaction, it moves back to the right (left). This scattering process may be 
depicted by the correlation function of $\langle J_L (\t, \s) J_R(\t^\prime,\sp) \rangle$ or 
$\langle J_R (\t, \s) J_L(\t^\prime,\sp) \rangle$. (See Fig. 1.) We expect that the correlation functions
$\langle J_L (\t, \s) J_L(\t^\prime,\sp) \rangle$ and $\langle J_R (\t, \s) J_R(\t^\prime,\sp) \rangle$ 
are not affected by the contact tunneling interaction. Let us calculate those correlation functions of the 
current operators. 

\begin{figure}[htbp]
   \begin {center}
    \epsfxsize=0.7\hsize
%
	\epsfbox{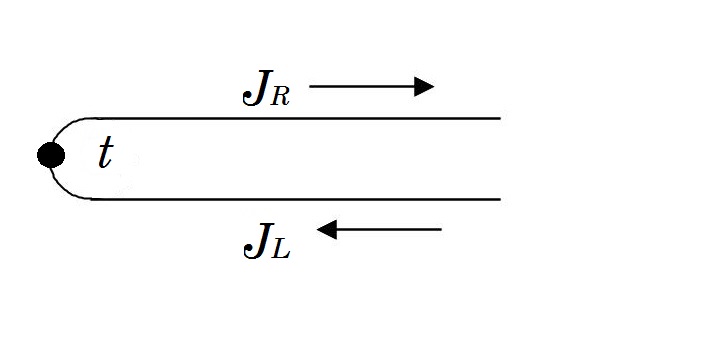}
   \end {center}
   \caption {\label{cond} Current Operators on the Half Line}
\end{figure}

The current operators are written in terms of the oscillator operators as 
\beq
J_L(\s) = \sum_{n, m} \psi^\dag_n \psi_m e^{-i(n+m)\s}, ~~ J_R(\s) = \sum_{n, m}
\tilde\psi^\dag_n \tilde\psi_m e^{i(n+m)\s},
\eeq
where $n, m \in {\bf Z}+ 1/2$. The correlation function of the two current operators $J_L$ on the boundary 
is given by 
\beq
\langle 0 \vert J_L (\s) J_L(\s^{\prime}) \vert B \rangle 
&=& \sum_{p,q,r,s \in {\bf Z} +1/2}
\langle 0 \vert  :\psi^\dag_p \psi_q : :\psi^\dag_r \psi_s :  
\prod_{n=1/2}^\infty:\exp\left\{-\frac{t^2}{n} \psi^\dag_{-n} \psi_n \right\} \nn\\
&&\exp \left\{-\frac{t^2}{n} \tilde\psi_{-n} \tilde\psi_n^\dag \right\}: 
\exp\left\{i\left(1-\frac{t^2}{n}\right) \psi^\dag_{-n} \tilde\psi^\dag_{-n}
\right\} \nn\\
&&\exp\left\{i\left(1-\frac{t^2}{n}\right) \psi_{-n} \tilde\psi_{-n} \right\}\vert 0 \rangle 
e^{-i(p+q)\s} e^{-i(r+s)\s^\prime} \nn\\
&=&\sum_{p,q,r,s \in {\bf Z} +1/2} \langle 0 \vert  :\psi^\dag_p \psi_q : :\psi^\dag_r \psi_s :
\vert 0\rangle e^{-i(p+q)\s} e^{-i(r+s)\s^\prime}.
\eeq
The vacuum on the LHS is annihilated unless $p$ and $q$ are positive half-odd integers. 
Since the current operators do not 
have the right movers, we should drop the terms containing the right movers.
The only non-vanishing components are those with $r= -p$, $s=-q$. The current correlation function 
$\langle J_LJ_L \rangle$ is not affected by the point contact interaction as expected. Thus,
\beq \label{JLJL}
\langle 0 \vert J_L (\s) J_L(\s^{\prime}) \vert B \rangle &=& \sum_{p, q=1/2}^\infty e^{-ip(\s-\s^\prime)}
 e^{-iq(\s-\s^\prime)} \nn\\
 &=& \frac{z z^\prime}{(z-z^\prime)^2}.
\eeq
where $z = e^{i\s}$, $z^\prime = e^{i\s^\prime}$. The correlation function of the current operators
$\langle 0 \vert J_L (\t,\s) J_L(\t^\prime,\s^{\prime}) \vert B \rangle$ is obtained from Eq.(\ref{JLJL})
$\langle 0 \vert J_L (\s) J_L(\s^{\prime}) \vert B \rangle$ simply by replacing $z$ and $z^\prime$ as
\beq
z = e^{i\s} \rightarrow z = e^{\t+ i\s}, ~~ z^\prime = e^{i\s^\prime} \rightarrow 
z^\prime = e^{\t^\prime+ i\s^\prime} .
\eeq

A similar procedure works also for 
$\langle 0 \vert J_R (\s) J_R(\s^{\prime}) \vert B \rangle$
\beq
\langle 0 \vert J_R (\s) J_R(\s^{\prime}) \vert B \rangle 
&=& \sum_{p,q,r,s}
\langle 0 \vert  :\tilde\psi^\dag_p \tilde\psi_q : :\tilde\psi^\dag_r \tilde\psi_s :  
\prod_{n=1/2}:\exp\left\{-\frac{t^2}{n} \psi^\dag_{-n} \psi_n \right\} \nn\\
&&\exp \left\{-\frac{t^2}{n} \tilde\psi_{-n} \tilde\psi_n^\dag \right\}: 
\exp\left\{i\left(1-\frac{t^2}{n}\right) \psi^\dag_{-n} \tilde\psi^\dag_{-n}
\right\} \nn\\
&&\exp\left\{i\left(1-\frac{t^2}{n}\right) \psi_{-n} \tilde\psi_{-n} \right\}\vert 0 \rangle 
e^{i(p+q)\s} e^{i(r+s)\s^\prime} \nn \\
&=&\sum_{p,q,r,s} \langle 0 \vert  :\tilde\psi^\dag_p \tilde\psi_q : :\tilde\psi^\dag_r \tilde\psi_s :
\vert 0\rangle e^{i(p+q)\s} e^{i(r+s)\s^\prime} \nn\\
&=& \sum_{p=1/2,\, q=1/2}^\infty e^{ip(\s-\s^\prime)} e^{iq(\s-\s^\prime)} \nn \\
&=& \frac{\barz \barz^\prime}{(\barz-\barz^\prime)^2}
\eeq
where $\barz = e^{-i\s}$, $\barz^\prime = e^{-i\sp}$. The current correlation function 
$\langle J_R J_R \rangle$ is also unaffected by the point contact interaction as expected.

The only non-trivial correlation function is $\langle J_L J_R \rangle$, which is given by
\beq
\langle 0\vert  J_L (\s) J_R(\s^\prime) \vert B \rangle 
&=&\sum_{p,q,r,s}
\langle 0 \vert  :\psi^\dag_p \psi_q : :\tilde\psi^\dag_r \tilde\psi_s : 
\vert B \rangle e^{-i(p+q)\s}e^{i(r+s)\s^\prime} \nn\\
&=& \sum_{p,q,r,s}
\langle 0 \vert  :\psi^\dag_p \psi_q : :\tilde\psi^\dag_r \tilde\psi_s : \prod_{n=1/2}^\infty:\exp\left\{-\frac{t^2}{n} \psi^\dag_{-n} \psi_n \right\} \nn\\
&&\exp \left\{-\frac{t^2}{n} \tilde\psi_{-n} \tilde\psi_n^\dag \right\}: 
\exp\left\{i\left(1-\frac{t^2}{n}\right) \psi^\dag_{-n} \tilde\psi^\dag_{-n}
\right\} \nn\\
&&\exp\left\{i\left(1-\frac{t^2}{n}\right) \psi_{-n} \tilde\psi_{-n} \right\}\vert 0 \rangle 
e^{i(p+q)\s} e^{i(r+s)\s^\prime} .
\eeq
By some algebra, we find 
\beq \label{JJ}
\langle 0\vert  J_L (\s) J_R(\s^\prime) \vert B \rangle 
&=& \sum_{p,\, q =1/2}^\infty \left(1 - \frac{t^2}{p}\right) \left(1 - \frac{t^2}{q}\right) e^{-i(p+q)(\s-\s^\prime)} \nn\\
&=& \left(\sum_{p=1/2}^\infty \left(1-\frac{t^2}{p}\right)e^{-ip(\s-\s^\prime)}\right)^2 . 
\eeq
We may define the mode (frequency) dependent conductance $G_{pq}$ as 
\begin{subequations}
\beq
\langle 0\vert  J_L (\t,\s) J_R(\t^\prime,\s^\prime) \vert B \rangle 
&=& \sum_{p,\, q =1/2}^\infty\left(G_{pq} +1\right) (z\barz^\prime)^{-p} (z\barz^\prime)^{-q}, \\
G_{pq} &=& \left(1 - \frac{t^2}{p}\right) \left(1 - \frac{t^2}{q}\right) -1 \label{conduct2}
\eeq
\end{subequations}
where  $z=e^{\t+i\s}$ and $\barz^\prime = e^{\t^\prime-i\sp}$.
The frequency dependency of the conductance is due to the non-local nature of the 
contact tunneling interaction. The contact interaction acts on the lower frequency modes more strongly.
The high frequency modes are affected only weakly by the contact tunneling interaction . 
For small $t$, $G_{pq} <0$. Eq.(\ref{conduct2}) indicates that the interaction suppresses the 
conductance and its effect decreases as the frequency increases.
Using the following series expansion
\beq
\ln\left(\frac{1+x}{1-x}\right) = 2 \left( x + \frac{x^3}{3} + \frac{x^5}{5} + \frac{x^7}{7}+ \cdots \right),
\eeq
we easily obtain the current correlation function 
$\langle 0\vert  J_L (\t,\s) J_R(\t^\prime,\s^\prime) \vert B \rangle$ in a closed form 
\beq \label{JLJR2}
\langle 0\vert  J_L (\t,\s) J_R(\t^\prime,\s^\prime) \vert B \rangle = 
\left( \frac{\sqrt{z {\bar z}^\prime}}{z {\bar z}^\prime-1} - 2 t^2 \ln \left( \frac{\sqrt{z {\bar z}^\prime}+1}{\sqrt{z {\bar z}^\prime}-1}\right)\right)^2,
\eeq
where $z=e^{\t+i\s}$ and $\barz^\prime = e^{\t^\prime-i\sp}$.

The correlation function as given by Eq.(\ref{JLJR2}) is valid only when $t$ is small, 
since we define the boundary state $\vert B\rangle$ as a perturbative state
from the Neumann state, correpsonding to the state with $t=0$.
As $t$ increases the point-contact tunneling interaction becomes strong. But for a large value of $t$ we 
cannot simply extrapolate the correlation function Eq.(\ref{JJ}) and Eq.(\ref{JLJR2})
which are valid only for small $t$. In the large $t$ limit, the boundary state turns into the 
Dirichlet state Eq.(\ref{Dirichleta},\ref{Dirichletb}). In this strong coupling region, we need to redefine the perturbation theory, 
around the Dirichlet state. It may accomplished by the T-dual transformation of string theory. 

\section{Conclusions}

In this paper we discuss the model of resonant point-contact tunneling of a single Luttinger-liquid lead,
which is the simplest model of this type. The model is initially defined on an infinite line with a single 
chiral Fermion and a mid-point contact interaction. In order to apply the boundary state formulation of 
string theory, we recast the model in the unfolded setup into the folded setup. In the folded setup 
the model is described by a Dirac spinor with two components. The Fermi field operators, 
corresponding to the creation and annihilation operators of charge on the resonant state, are 
also extended to a spinor with two components. Integrating out this Fermi field yields the boundary 
effective action, which is nonlocal. Since the effective boundary action is only quadratic in the field operator, 
the boundary state can be explicitly constructed. The constructed boundary state is shown to reproduce correctly 
the continuity condition of the model in the unfolded setup as boundary condition. 

Correlation functions of the physical operators can be calculated by using the explicit expression of the 
boundary state. As an immediate application the correlation functions of current operators are evaluated.
Despite the simple structure of the model,
it shares the intriguing features of the fully fledged models. The point contact interaction acts on the 
lower frequency modes more strongly. The high frequency modes are only weakly affected by the interaction. 
The folding procedure and the boundary state formulation are certainly applicable to more complex models, 
which include the resonant multilead point-contact tunneling \cite{Nayak}, the quantum Brownian motion 
on a honeycomb lattice \cite{Nayak,chamon,Lee2009q}. 
Extensions of this work along this direction will be given somewhere else. 
\vskip 1cm

\noindent{\bf Acknowledgments}

This work was supported by Kangwon National University.


\begin{thebibliography}{0}

\bibitem{Nayak}  C. Nayak, M. P. A. Fisher, A. W. W. Ludwig and H. H. Lin, 
{\it Resonant multilead point-contact tunneling}, 
Phys. Rev. B {\bf 59}, 15 694 (1999).


\bibitem{Luttinger}
J. M. Luttinger, 
{\it An exactly soluable model of a many-fermion system}, J Math Phys {\bf 4}, 9
(1963)

\bibitem{Haldane} F. D. M. Haldane, 
{\it Luttinger liquid theory’ of one-dimension quantum fluids: I. Properties
of the Luttinger model and their extension to the general 1D interacting spinless Fermi
gas}, J Phys C: Solid State Phys {\bf 14}, 2585 (1981)


\bibitem{Hewson} A. C. Hewson, 
{\it The Kondo Problem to Heavy Fermions}, 
Cambridge University Press (1997).

\bibitem{Saleur1998} H. Saleur, 
{\it Lectures on Non Perturbative Field Theory and Quantum Impurity Problems},
arXiv:cond-mat/9812110v1 (1998)

\bibitem{Saleur2000} H. Saleur, 
{\it Lectures on Non Perturbative Field Theory and Quantum Impurity Problems: Part II},
arXiv:cond-mat/0007309


\bibitem{Affleck:1990by}
I.~Affleck and A.~W.~W.~Ludwig,
{\it The Kondo Effect, Conformal Field Theory And Fusion Rules},
Nucl.\ Phys.\ B {\bf 352}, 849 (1991).

\bibitem{fendley1995a} P. Fendley, A. W. W. Ludwig and H. Saleur,
Phys. Rev. Lett. {\bf 74} 3005 (1995).

\bibitem{fendley1995} P. Fendley, A. W. W. Ludwig and H. Saleur,
{\it Exact nonequilibrium transport through point contacts in quantum wires and fractional quantum Hall devices},
Phys. Rev. B {\bf 52} (1995) 8934. [arXiv:cond-mat/9503172]

\bibitem{wong} E. Wong and I. Affleck,
{\it Tunneling in Quantum Wires: a Boundary Conformal Field Theory Approach},
arXiv:cond-mat/9311040v1 (1993).
Nucl. Phys. B {\bf 417}, 403 (1994). 

\bibitem{Affleck1994}
A. Affleck and W. W, Ludwig, J. Phys. A {\bf 27}, 5375 (1994).

\bibitem{callan90} 
C G. Callan, Jr. and L. Thorlacius, 
{\it Open String Theory as Dissipative Quantum Mechanics},
Nucl. Phys. B {\bf 329}, 117 (1990).

\bibitem{callan91}
C. G. Callan, Jr. and D. Freed,
{\it Phase Diagram of the Dissipative Hofstadter Model},
Nucl. Phys. B {\bf 374}, 543 (1992).
 [hep-th/9110046]. 

\bibitem{Callan1994} C. G. Callan, 
{\it Exact $c=1$ Boundary Conformal Field Theories}, 
Phys. Rev. Lett. {\bf 72}, 1968 (1994). 

\bibitem{Lee:2005ge}
T.~Lee and G.~W.~Semenoff,
{\it Fermion representation of the rolling tachyon boundary conformal field theory}, 
JHEP {\bf 0505}, 072 (2005).

\bibitem{Hassel} 
M. Hasselfield, T. Lee, G. W. Semenoff, P. C. E. Stamp,
{\it Critical Boundary Sine-Gordon Revisited} 
Ann. Phys. {\bf 321}, 2849 (2006).
[hep-th/0512219]

\bibitem{Lee:06}
 T.~Lee,
 ``The final fate of the rolling tachyon,"
JHEP {\bf 0611}, 056 (2006).

\bibitem{Lee2008}
T. Lee,
``Applications of Thirring model to inhomogeneous rolling tachyon and dissipative quantum
 mechanics,"'
JHEP {\bf 02} 090 (2008).

\bibitem{Lee2009}
T. Lee,
{\it String Theory and Dualities in the Quantum Dissipative Hofstadter System}
Int. J. Mod. Phys. {\bf A24}, 6141 (2009).


\bibitem{brown} 
L. S. Brown,
{\it Quantum Field Theory}, Cambridge University Press (1992).


\bibitem{chamon}
C.~Chamon, M.~Oshikawa, and I.~Affleck,
{\it Junctions of three quantum wires and the Dissipative Hofstadter Model},
Phys. Rev. Lett.  {\bf 91}, 206403 (2003).
[cond-mat/0509675].


\bibitem{Lee2009q}
T. Lee, 
{\it Quantum Brownian Motion on a Triangular Lattice and Fermi-Bose Equivalence:
An Application of Boundary State Formulation},
JHEP {\bf 03} 078 (2009).














\end{thebibliography}
\end{document}